Review

# The s- and r- components of the proto-solar composition

**N. Prantzos**[1,a], **S. Cristallo**[2,3,b], **C. Abia**[4,c]

[1] Institut d'Astrophysique de Paris, CNRS and Sorbonne Université, 98bis bd Arago, 75014 Paris, France
[2] INAF-Osservatorio Astronomico d'Abruzzo, Via Maggini snc, 64100 Teramo, Italy
[3] INFN-Sezione di Perugia, Via A. Pascoli snc, 06121 Perugia, Italy
[4] Dept. Física Teórica y del Cosmos, Universidad de Granada, Avda. Fuentenueva s/n, 18071 Granada, Spain

**Abstract** We present a brief overview of the various methods proposed to derive the s- and r-components of the proto-solar chemical composition and we discuss some recent developments in the field, including the impact of rotating massive stars, nuclear measurements, physics of low mass asymptotic giant branch stars, isotopic composition of presolar SiC grains and new derivations.

## 1 Introduction

The decomposition of the Solar system abundances of heavy elements into their s- and r-components played, and will continue to play, a pivotal role in our understanding of the underlying nuclear processes and the physics and evolution of the corresponding sites. The s-contribution can be more easily determined, since isotopes dominated by the s-process form close to the $\beta$-stability valley. Their nuclear properties ($\beta$-decay half-times, nuclear cross-sections, etc.) are quite easily measured and the astrophysical hosting sites are sufficiently understood today. However, due to the large astrophysical and nuclear physics uncertainties related with the r-process, its contribution to the isotopic solar abundances has been deduced mainly by a simple subtraction of the s-process contribution from the observed solar value (the so-called residual method, [1–3]).

Roberto Gallino played an important role in this field by extensively studying the astrophysical sites of the s-process (He-burning cores of massive stars and He-burning shells of AGBs) and identifying key uncertainties (both nuclear and stellar) that impact the distribution of the produced heavy nuclei [3–6]. In this paper – a short scientific tribute to his memory – we present a brief overview of the various methods that have been used to evaluate the s- and r-contributions to the solar system composition (Sect. 2) and discuss some recent developments in that field (Sect. 3).

## 2 Determination of s- and r- abundances

### 2.1 The canonical model

The "classical" (or "canonical") s-process model was originally proposed by Burbidge et al. [7] and developed by Clayton and Rassbach [8]. In this model two main assumptions are made: (a) the temperature of the s-process is constant, allowing one to adopt well determined neutron capture cross sections; (b) the nuclei on the s-process path are stable, that is their $\beta$-decay timescales are much larger than their neutron-capture ones ($\tau_\beta >> \tau_n$) or sufficiently short-lived that the neutron capture chain continues with the daughter nucleus; ($\tau_\beta << \tau_n$). However, this second assumption is not valid on the branches of the s-process ($\tau_\beta \sim \tau_n$), which require special treatment [9]. In addition, the classical model assumes that some stellar material composed of iron nuclei is exposed only to the superposition of three exponential distributions of the time-integrated neutron exposure, defined as $\tau_o = \int_o^t N_n v_T dt$ (where $v_T$ is the thermal neutron velocity at temperature T). These three exponential distributions are usually referred to as the "weak" component (responsible for the production of the $70 \leq A \leq 90$ s-nuclei), the "main" component (for the $90 \leq A \leq 204$ isotopes) and the "strong component" (for $A > 204$). For long-enough exposures, the equations governing the evolution of the s-nuclei abundances

N Prantzos, S Cristallo and C Abia contributed equally to this work.

[a] e-mail: prantzos@iap.fr (corresponding author)
[b] e-mail: sergio.cristallo@inaf.it
[c] e-mail: cabia@ugr.es

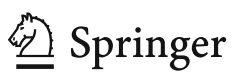

result in equilibrium between the production and destruction terms, leading to a constant, product $\sigma_A N_A$, of the neutron capture cross section and the abundance of the s-process isotope. Although this condition is not reached completely, the product $\sigma_A N_A$ shows a very smooth dependence on the mass number [10]. Therefore, the product $\sigma_A N_A$ for a given isotope is fully determined by the cross section, after the parameters $\tau_o$ and the number of neutrons captured per $^{56}$Fe seed nucleus are fixed. The goal of the classical approach is to fix the empirical values $\sigma_A N_A$ for the s-only isotopes, i.e., nuclei that are shielded against the r-process by the corresponding stable isobar with charge $Z-1$ or $Z-2$. Once the Solar system s-only distribution is fitted, the s-contribution for the rest of the "mixed" isotopes (with both a s- and r-contribution) is automatically obtained. Finally, the r-contribution is derived just by subtracting this s-contribution $\mathrm{N}_{s,A}$ from the total Solar system abundance measured $\mathrm{N}_A$. This classical method has been used frequently in the literature, providing satisfactory results as the measurement of neutron cross sections has been improving over the years [9,11,12].

### 2.2 The "multi-event model"

The classical model is affected not only by observational and nuclear input data uncertainties, but also by the assumption that the s-process operates at a fixed constant temperature and neutron and electron densities, and by the hypothesis that the irradiation can be considered as an exponential one.

To test the influence of these assumptions, [13,14] developed the so-called "multi-event" s-process, which constitutes a step forward in the canonical method. The multi-event approach assumes a superposition of a number of canonical events taken place in different thermodynamic conditions, namely: a temperature range $1.5 \leq \mathrm{T(K)}/10^8 \leq 4$, neutron densities $7.5 \leq \log \mathrm{N_n}(\mathrm{cm}^{-3}) \leq 10$ and a unique electron density $\mathrm{N_e} = 10^{27}$ cm$^{-3}$. Each canonical event is characterised by a given neutron irradiation in the $^{56}$Fe seed nuclei for a given time at a constant temperature and neutron density. These conditions try to mimic the astrophysical conditions characteristic of the site of the s-process, although it is well known that temperature and neutron density are not constant during the s-process [6,15]. The s-only nuclei abundance distribution obtained with that method is remarkably close to the Solar observed one because of the minimisation procedure adopted in the selection of the aforementioned parameters. However, it presents non-negligible deviations from the classical method in the regions $A \leq 90$ and $A \geq 204$, mainly because the resulting neutron exposures in the multi-event model clearly deviate from exponential.

Within the multi-event model it was possible to evaluate the major uncertainties (both nuclear and due to abundance measurements) affecting the prediction of the s-(r-)abundance distribution. Goriely [13] concluded that the uncertainties in the observed meteoritic abundances and the relevant $(n, \gamma)$ rates have a significant impact on the predicted s-component of the Solar abundance and consequently on the derived r-abundances, especially concerning the s-dominated nuclei (see also [16,17]).

### 2.3 The "stellar" model

Stellar models of low- and intermediate mass (LIM) stars during the AGB phase and of massive stars during hydrostatic core He-burning and shell C-burning (the two widely recognised sites of the s-process) have shown that the interplay of the different thermal conditions for the $^{13}$C and $^{22}$Ne neutron sources is hardly represented by a single set of effective parameters constant in time [18–20], such as those used in the classical (or the multi-event) approach. In an effort to overcome this shortcoming, the results of the "*stellar*" model have been used to estimate the contributions of the s- and r-process to the Solar system abundances. This method is based on post-processing nucleosynthesis calculations performed in the framework of realistic stellar models. The first attempt to apply this method was made by [3,5], and more recently by [21] with updated nuclear input. These authors showed that the Solar s-process main component can be reasonably reproduced by a post-processing calculation from a particular choice (mass and extension) of the $^{13}$C pocket (the main neutron source in AGB stars) by averaging the results of stellar AGB models [5] between 1.5 and 3 $\mathrm{M}_\odot$ with [Fe/H]$\sim -0.3$. This model is particularly successful in reproducing the solar abundances of the s-only nuclei and showed general improvements with respect to the classical method, especially in the mass region $A < 88$ [3]. In fact, all of these nuclei (mainly produced by the weak s-component) are synthesised in much smaller quantities. This difference is caused by the very high neutron exposures reached in the stellar model, which favour the production of heavier elements. In particular, on the s-termination path, $^{208}$Pb is produced four times more than in the classical approach. Nevertheless, the stellar model used to derive the physical inputs of post-process calculations are affected by several theoretical uncertainties. One of the less constrained physical mechanisms is the one that leads to the formation of the $^{13}$C pocket, which forms at the base of the convective envelope after each third episode of dredge-up [6].

Different processes have been proposed as responsible for the formation of such a pocket: convective overshoot [22], gravity waves [23,24], opacity induced overshoot [25] and mixing induced by magnetic mixing [26]. Other critical quantities are the mass fraction dredged-up after each thermal instability (third dredge up, TDU) during the AGB phase, and the mass-loss rate. Actually, the two processes are degenerate, since the number (and efficiency) of TDUs is determined by the mass of the H-exhausted core and of the H-rich convec-tive envelope, which in turn depend on the adopted mass-loss rate. However, AGB stellar models show that an asymptotic s-process distribution is reached after a limited number of pulses, so the mass-loss uncertainty mainly affects the total yield of the s-processed material and not so much the shape of the resulting distribution [27,28].

In the stellar model, the r-residuals are calculated subtracting the arithmetic average of the 1.5 and 3 $M_\odot$ models with [Fe/H]$\sim -0.3$ ($Z \sim 0.5 \times Z_\odot$)[1] best reproducing the main s-component to the observed Solar abundances. In Arlandini et al. [3] the s- and r-components obtained by the stellar model method are compared to the classical one for nuclei $A > 88$, together with the corresponding uncertainty determined from the cross sections and Solar abundances. However, uncertainties in the s- and r-residuals coming from the stellar model itself are difficult to estimate.

The massive star contribution to the solar s- only composition has been explored with non-rotating stellar models in e.g. [29,30] and more recently with rotating massive stars in [20,31–34]. Such models have their own uncertainties (mass loss, mixing, nuclear inputs, etc). The role of rotation, in particular, is poorly explored and understood at present. The main reason is that the rotation driven instabilities are included in a parametric way, and this means that the efficiency with which fresh protons are ingested in the He-burning zone is not based on first principles but it is determined by two free parameters that must be calibrated. A detailed discussion of the calibration adopted in some rotational models can be read in [20]. Moreover, since the proton ingestion scales directly with the initial rotational velocity (and hence the neutron flux as well), the adopted initial distribution of rotational velocities (IDROV) plays a pivotal role: in [35] it was shown that at least the average rotational velocity of the stars must be limited to $< 50$ km/s at metallicities [Fe/H]$> -1$, in order to avoid an overproduction of heavy nuclei, mainly in the Ba peak. But there are also other subtle indirect factors that may change the yields predicted by rotating models: in order to bring protons in an He active environment, at least part of the H rich mantle must be present while He is burning. A substantial change in the mass loss rate (e.g. due to the inclusion of a dust driven component to the mass loss rate or to the overcome of the Eddington luminosity) may affect the range of masses that retain a substantial fraction of the H rich mantle while the stars are in the central He-burning phase. Finally, nuclear uncertainties, like the one still affecting the $^{22}$Ne($\alpha$, n) reaction, may play a major role (see Sect. 3.1).

[1] We adopt here the usual notation [X/H]= log $(X/H)_\star$ − log $(X/H)_\odot$, where $(X/H)_\star$ is the abundance by number of the element X in the corresponding object.

### 2.4 The "galactic" model

The uncertainty of stellar models is one of the reasons why the validity of the stellar method has been questioned [14]. Another one is that this method does not consider the solar s-(r-)process abundance distribution in an astrophysical framework, i.e. as the result of all the previous generations of stars which polluted the interstellar medium prior to the formation of the Solar system. In particular, these generations of stars covered a wide range of metallicities and *not a unique value (or even a limited range of values)* of [Fe/H] as is assumed in the classical and stellar methods. For example, it is well known that at low metallicities a large neutron/seed ratio is obtained, leading to the production of the heaviest s-nuclei, while at high metallicities the opposite occurs [4,36].

Ultimately, the abundances of the solar system s-(r-) processes have to be understood in the framework of a galactic chemical evolution (GCE) model. This is certainly a difficult task that requires a good understanding of the star formation history in the various regions of the Galaxy, stellar evolution, and the interplay between stars and interstellar gas, among other things. We are still far from fully understanding these issues. Therefore, this third method is based on a necessarily schematic description of the situation considering the chemical evolution of our Galaxy, accounting for the fact that the site(s) of the r-process have not been clearly identified yet. Attempts to obtain the s- and r-components of the solar composition from a GCE model were pioneered by [36], later updated by [37–39]. These authors employed a GCE code that adopts s-process yields from AGB stellar models by [5] in a range of masses and metallicities. Regarding the r-process yields, and for elements from Ba to Pb, they estimated the contribution to the Solar system by subtracting the s-residuals from the Solar abundances. Then, they scaled the r-process yields to the yield of a primary element mainly produced in core collapse supernovae, which they assumed to occur in the mass range 8–10 $M_\odot$. They derived the contribution of the weak s-process from [30]. On the other hand, for the lighter elements, in particular for Sr–Y–Zr, they deduced the r-residuals and, thus, the r-process yields, from the abundance pattern found in CS 22892-052 [40], by assuming that the abundance signatures of this star is of *pure r-process origin* (i.e., any contamination by other possible stellar sources is hidden by the r-process abundances).

The model of [38] resulted in good agreement with the Solar s-only isotopic abundances between $^{134,136}$Ba and $^{204}$Pb, also showing that the solar abundance of $^{208}$Pb is well reproduced by metal-poor AGB stars, without requiring the existence of a "strong" component in the s-process as is done in the classical method. However, below the magic number $N = 82$, they found a significant discrepancy between the abundance distribution obtained with their GCE model and the Solar system values. It turned out that their GCE model under-produces the solar s-process component of the abundances of Sr, Y and Zr by $\sim 20\% - 30\%$ and also the s-only isotopes from $^{96}$Mo up to $^{130}$Xe. They basically confirm the results by [36], who postulate the existence of another source of neutron-capture nucleosynthesis called the *light element primary process (LEPP)*. They argued that this process is different from the s-process in AGB stars and also different from the weak s-process component occurring in massive stars. The updates of this study by [38,39], reach the same conclusion.[2] In particular, these two studies ascribe a fraction ranging from 8% to 18% of the solar abundances of Sr, Y, and Zr to this LEPP, and suggest that lighter elements from Cu to Kr could also be affected. On the other hand, they obtained an r-process fraction at the Solar system ranging from 8% (Y) to 50% (Ru). In fact, [36] also proposed the existence of another LEPP to explain the high [Sr/Ba] ratios observed in very metal-poor stars. The nature and nucleosynthesis mechanism of this second extra source of the light trans-iron elements remains debated. However, it is generally believed to be linked to massive stars, as they evolved rapidly and began to enrich the interstellar gas in the earliest phases of Galactic chemical evolution. However, very recently, based on the Ba even/odd isotopic ratios measured in metal-poor stars, [41] concluded that this early LEPP is a non-standard s-process producing Sr but also Ba associated with massive stars [20,34].

The need for a solar LEPP has been questioned by [27] and later by [26] on the basis of a simple GCE model using updated s-process yields from AGB stars [42] and AGB stellar models only, respectively. These studies show that a fraction of the order of that ascribed to the LEPP in the predicted solar abundances of Sr, Y and Zr can be easily obtained, for instance, by just a moderate change in the star formation rate prescription in a GCE model, still fulfilling the main observational constrains in the solar neighbourhood. The same effect can be found by modifying the AGB stellar yields as a result of nuclear uncertainties or by choosing a different mass and profile of the $^{13}$C pocket. Introducing such changes in the GCE models (i.e., stellar yields), one can easily account for the missing fractions of the solar abundance of these elements within the observational uncertainties.

### 2.5 The "iterative galactic" model

The LEPP hypothesis was also questioned in [35] who used metallicity-dependent s-process yields from rotating massive stars (i.e. the "weak" s-process) in a GCE model. The stellar yields adopted in that paper were from an extended grid of stellar masses, metallicities, and rotation velocities from [20]. The model satisfies all the main observational constraints that can be expected from a 1-zone model for the solar neighbourhood. In particular, the isotopic distribution of nuclei up to the Fe-peak at Solar system formation (4.5 Gyr ago) is well reproduced. Subsequently, the same group used the GCE model of [35]) to derive the contributions of the s- and r-process to the solar isotopic abundances in the full mass range from $^{69}$Ga to $^{209}$Bi through a new method [43]. The method consists in running first a model with only the s-component for all heavy isotopes, i.e. without considering any r-component in the stellar ejecta (model M0). Then, the r-component is introduced (model M1) based on some prior estimate of it (in that case, the r-fractions estimated by Goriely [13] or Sneden et al. [11]). It is assumed that the r-process is primary in nature and follows the evolution of an alpha element, as suggested by observations of, e.g. [Eu/Fe] for the evolution of the Milky Way disc. The s-fractions of all nuclei $f_s$ are thus obtained as the results of M0/M1 at Solar system formation. These fractions are by construction $\leq 1$ (independently if the corresponding isotopes are overproduced with respect to their solar abundance) and are equal to one for the 30 heavy isotopes which are *predefined* as s-only. This is one of the advantages of that method because the corresponding r-fractions, defined as $f_r = 1 - f_s$, are also positive independently of whether the corresponding nuclei in Model 0 or 1 are overproduced or not.

The r- fractions differ slightly from their initial value (the prior adopted). In order to make the GCE model self-consistent, the new r-fractions are then injected into it, and the model is run again. The new isotopic composition fits better the solar one (as indicated by a simple $\chi^2$ test), a new set of r-contributions is derived as previously, and iterations are repeated with the new r-component until the results do not vary sensitively any more. The final decomposition of the Solar system isotopic abundance distribution into s- and r-components with this "iterative GCE", (IGCE) method is then *fully self-consistent with the GCE model and the adopted stellar yields*.

Comparison of the s- and r-distribution models obtained with the IGCE method with measured solar abundances shows excellent agreement, especially when uncertainties in the measured abundances are considered. The most important deviations concern the s-only nuclei, because their abundance stems directly from the adopted stellar yields and the IMF, and there is no possibility to modify that by an adjustable r-component, as done for all other mixed nuclei (s+r or pure r). Since the final abundances are driven by the s-component of Model M0, it is clear that the most important deviations are expected to be found in the regions where the s-process dominates, namely, the three s-peaks. This is encouraging since it implies that a better treatment of the s-process in both LIM stars and rotating massive stars will allow one to reduce further those deviations from the solar composition by applying the IGCE method.

[2] [38,39] mainly focus on the impact of the different $^{13}$C pocket choices in AGB stars and weak s-process yields from massive stars on the s-process residuals at the epoch of the Solar System formation. [39] included yields from massive stars considering the impact of rotation in a limited range of masses and metallicities according to the models by [32].

In the study of [43], the resulting $\sigma N_{s,A}$ curve displays the classical feature of $\sigma N_{s,A} \sim$ constant between magic neutron numbers, but shows an interesting difference with the classical study of [11]: nuclei lying near branching points, like $^{148}$Nd, $^{170}$Er and $^{192}$Os, receive a fairly small s-contribution in [11] but a considerably larger one (factors 2-4) in the case of [43]; the latter results are, in general, in better agreement with those of [13,38], probably because those studies explore a larger range of (and/or more realistic) physical conditions in stellar interiors than the classical study of [11].

Compared to works based on different methods [11,13, 38], good general agreement is found between the various studies, but also some discrepancies. The most important ones are found in the region of the weak s-process, namely below A = 90. In that region, [11] find that $^{76}$Ge and $^{82}$Se are r-only nuclei (in agreement with [13]), while [43] find a substantial contribution from the s-process in rotating massive stars (25% and 11%, respectively).

## 3 Some recent developments

Since the study of [43], several developments have been made in the field of the determination of the components s- and r- of the proto-solar composition. They concern the reference proto-solar abundances, new measurements of neutron-capture cross sections, the stellar evolution models of low mass stars and the corresponding s-process yields, and recent evaluations of the s- and r-components using different methods.

The new compilation of proto-solar isotopic abundances of [44] presents some differences with respect to [45], which was used in the study of [43]. The most important concern the upward reassessment of the CNO abundances, resulting in a metallicity of $Z_{\odot} = 0.0178$. Regarding heavy isotopes, the changes are minor, as can be seen in Fig. 1 (top panel), where we display the comparison of the new compilation of proto-solar isotopic abundances from [44] with those of [45]. The sole indium isotope, $^{127}$I (an r-isotope), increases by $\sim$ 45%,while $^{79}$Br, $^{81}$Br and $^{180}$Ta increase by $\sim$ 15% and the isotopes of Hg decrease by $\sim$ 15%. Modifications to all other heavy isotopes are less than 10%.

These modifications are expected to affect the determination of the s- and r-components slightly through any of the methods presented in the previous section. In Fig. 1 (bottom panel) we present the $\sigma_A N_A$ curve calculated with the new set of abundances $N_A$ and neutron capture cross sections $\sigma_A$ from the *Kadonis* database [46], taken at 30 keV for the "weak" s-process nuclides (A < 90) which are produced in core He-burning massive stars from the $^{22}$Ne($\alpha$, n) neutron source and at 10 keV for the heavier nuclei of the "main" s-process nuclides (A > 90) which are produced in shell He-burning low mass stars from the $^{13}$C($\alpha$, n) neutron source. A comparison with Fig. 4 of [43] shows that the relative variations among the various isotopes are quite small. We note that the values of the two plateaus (in the ranges 90 < A < 135 and 140 < A < 204) were lower in [43] than in Fig. 1 here, because cross sections at 30 keV were used for the full range of A in the former case. Our analysis displays similar results to those obtained in figures 14 and 15 in [44].

### 3.1 Nuclear physics uncertainties for the s-process

The AGB models used in [43] have been calculated with the FUNS code [25,47]. The results are available online on the FRUITY database [42,48]. Those AGB models include a full nuclear network (from hydrogen to lead-bismuth, at the end point of the s-process) which is directly coupled to the physical evolution of the structure. The adopted nuclear network is described in detail in the aforementioned papers. However, in recent years, new strong reactions (involving charged particles or neutrons) and weak rates (in particular, $\beta$-decays) have become available.

In Fig. 2 we present the comparison between two AGB models with the same mass (M=2 $M_{\odot}$) and metallicity ($Z$=0.01) but different networks: the reference model [43] and the most updated (label "*Network 2025*": [49,50]). In the upper panel, we report the surface element distributions as a function of atomic number, in the usual spectroscopic notation. In the lower panel, we plot the total yield ratios between the new and reference models as a function of their atomic mass for s-only isotopes and magic nuclei. Both isotopic distributions have previously been normalised to their corresponding yields of $^{150}$Sm. In such a way, we can highlight the effects of adopted nuclear inputs on the isotopic relative distribution, while in the upper panel we can focus on absolute production factors.

The large reduction in $^{140}$Ce is due to the new neutron capture cross section adopted measured in the n_TOF experiment [51], which results in an increase of more than 20% at the AGB temperatures of interest. This led to a decrease in $^{140}$Ce and an increase of heavier isotopes. In contrast to cerium, we highlight a net decrease in the production of $^{134}$Ba, i.e., the s-only isotope with the largest overproduction with respect to $^{150}$Sm in the distribution presented by [43]. The explanation of such a decrease lies in the new adopted $^{134}$Cs $\beta$-decay, whose half-life has recently been re-calculated [52,53] and demonstrated to be smaller than [54] (the reference rate) at temperatures T > 200 MK. Thus, $^{134}$Cs can capture more neutrons, by-passing the nucleosynthetic channel to $^{134}$Ba.

Another interesting feature is the increase of $^{96}$Mo with respect to its neighbouring isotopes, due to the proposed

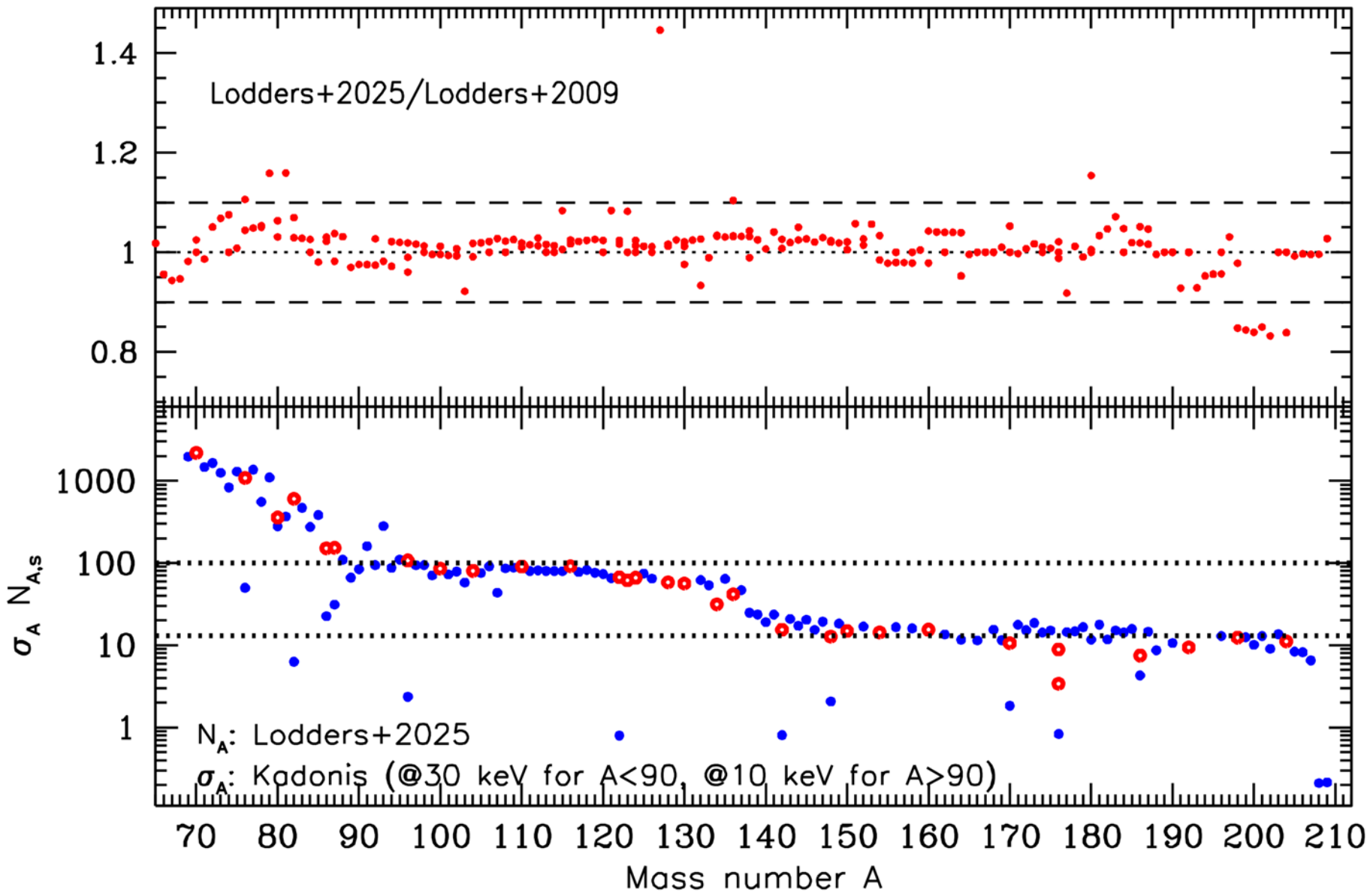

**Fig. 1** Top: comparison of heavy isotope abundances between the studies of [44] and [45]. Bottom: $\sigma_A N_{A,s}$ curve obtained from the proto-solar system isotopic abundances of [44], the s- fractions of [43] and the *Kadonis* data base of neutron capture cross-sections $\sigma_A$ (here taken at massive stars energies of 30 keV for the weak s-process for A < 90 and at 10 keV for the main s-process at A > 90). Red open symbols represent s-only nuclei

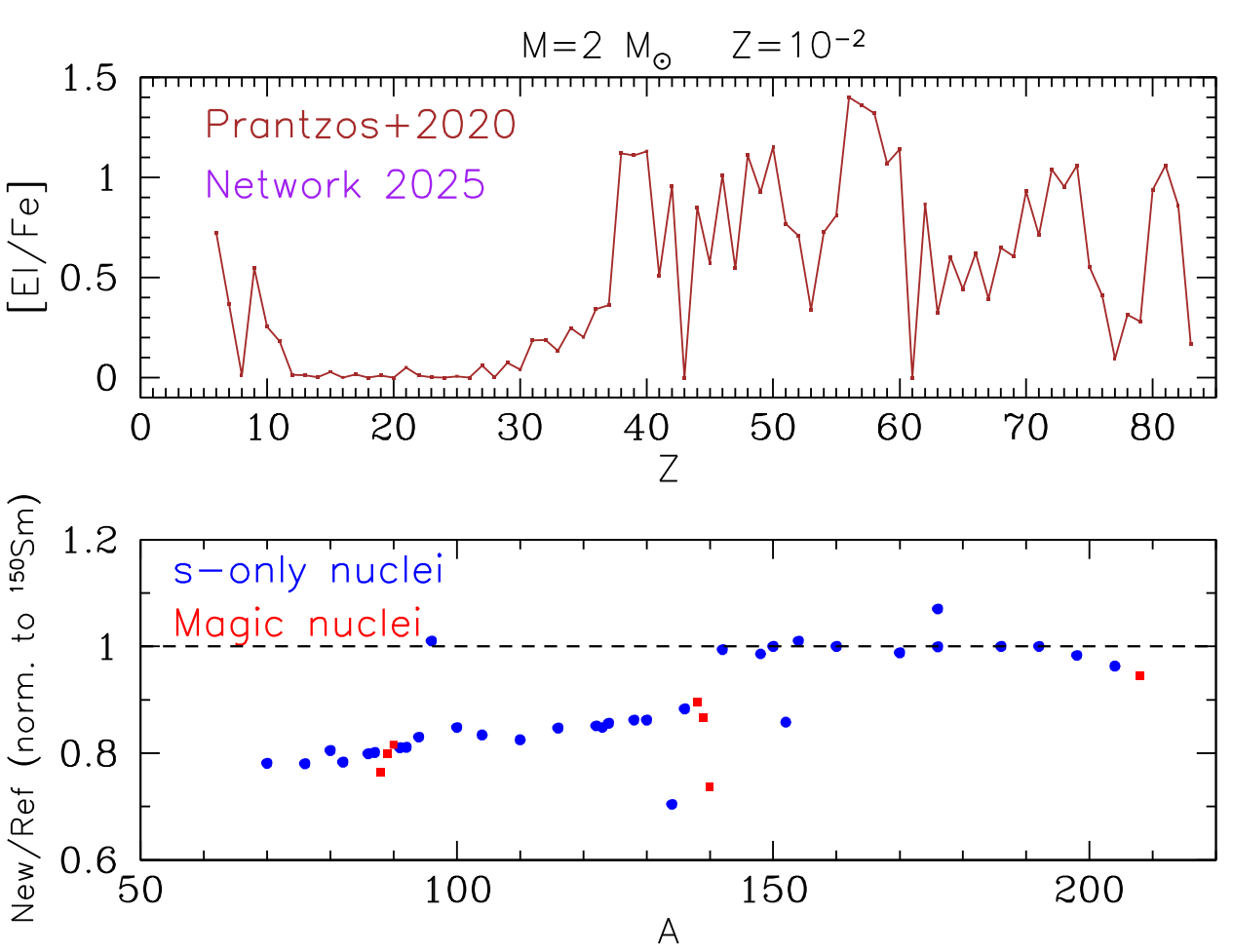

**Fig. 2** Comparison between the abundance enhancement (top panel) obtained with the 2 M$_\odot$, Z = 0.01 reference model [43] and a model computed with an updated nuclear network (see text for details)

increase of the nuclear rate of $^{95}$Mo(n, $\gamma$)$^{96}$Mo by [55]. In fact, this may represent a potential problem, considering the very good fit obtained by [43]. However, a new experimental campaign with the idea of measuring Mo isotope neutron capture cross sections took place at n_TOF and GELINA [56]: the publication of new experimental cross sections will be crucial to analyse this potential problem. Among the varied rates, we also implemented the cross sections governing the production of neutrons, i.e., the reactions $^{13}$C($\alpha$, n)$^{16}$O and $^{22}$Ne($\alpha$, $n$)$^{25}$Mg. The first reaction has been almost measured at its Gamow Peak [57] (i.e. the energy range of interest in the interior of AGB stars), and thus we do not expect important variations in the future. The situation of the $^{22}$Ne($\alpha$, n)$^{25}$Mg reaction, which provides a marginal neutron burst in AGB stars and represents the main neutron source in massive stars, is different. To date, there is a large disagreement between direct [58] and indirect experimental [59] measurements (up to a factor of 3 in the energies of interest). In our new AGB models, we adopt the rate proposed by [60], and in particular the evaluation of the cross sections not including indirect measurements. It is clear that a different choice (indirect vs direct) would imply large variations in the yields of the light s- isotopes, in particular those belonging to the weak s-process. This is also important when computing fast-rotating massive star models, in which the major neutron contributor is the $^{22}$Ne($\alpha$, n)$^{25}$Mg reaction. Nuclear uncertainties, added to those of the stellar models (treatment of convective borders, of rotation and of mass-loss rates), leave the resulting stellar yields on shaky ground.

### 3.2 Stellar evolutionary model uncertainties

A large source of uncertainty in AGB yields comes from the adopted mechanism to mimic the formation of the so-called $^{13}$C-pocket, i.e. the major neutron source in AGB stars [12, 18]. This requires the mix of a proper amount of hydrogen (not too little to have enough proton captures on $^{12}$C, not too much to avoid proton captures on $^{13}$C itself) beneath the inner border of the convective envelope. In AGB standard FRUITY models (i.e., those adopted in [43]), the pocket is formed by means of the opacity induced overshoot [25]. However, the rather poor agreement between FRUITY AGB models and laboratory isotopic heavy element measurements in presolar SiC grains [62], led [61] to implement a new type of mixing, triggered by the presence of magnetic fields. The inclusion of magnetic buoyancy triggers a non-convective mixing which attains a better match to isotopic ratios measured in presolar grains (in particular for strontium and barium).

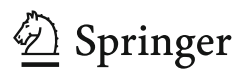

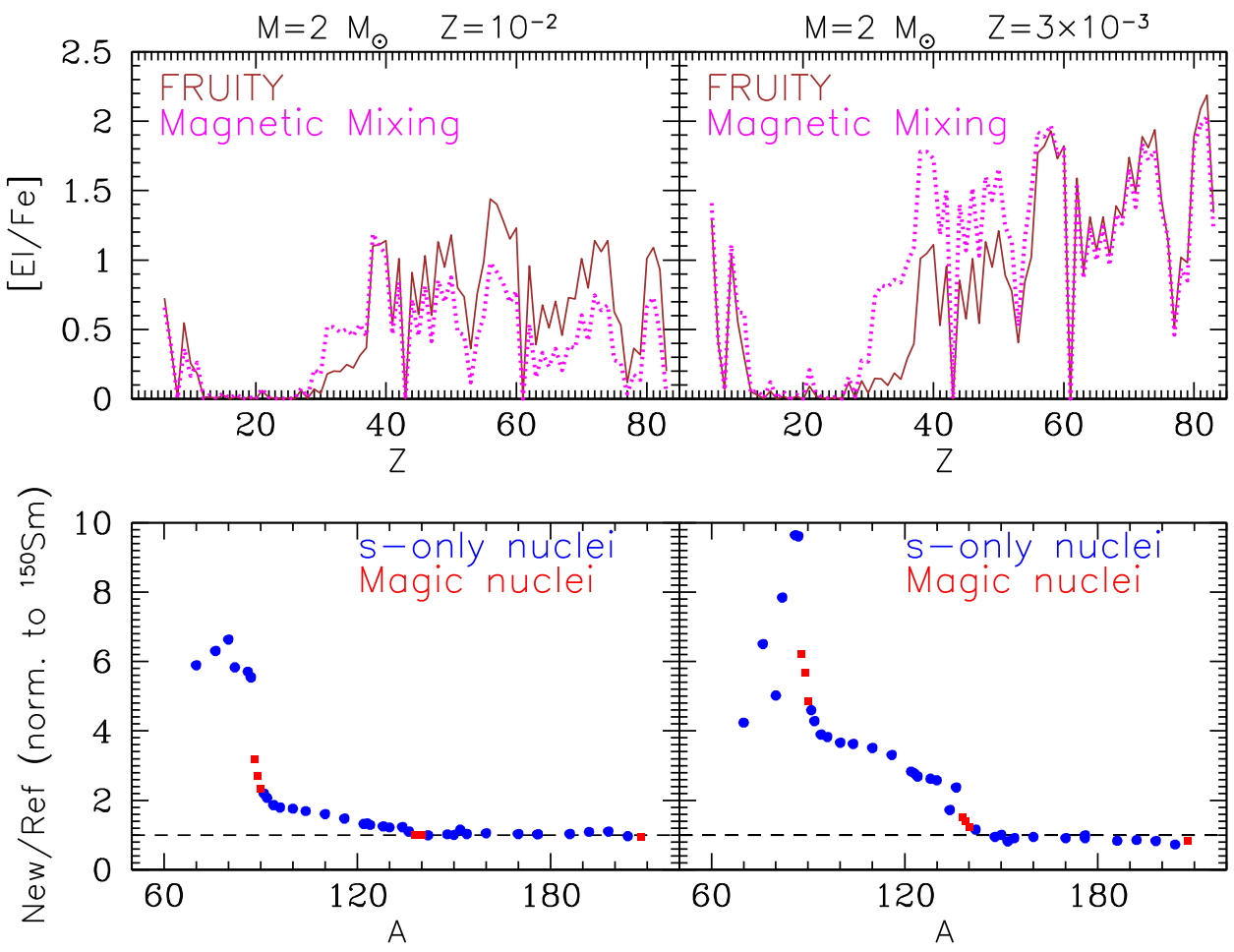

**Fig. 3** As in Fig. 2, but with a new prescription for the formation of the $^{13}$C pocket [61]. See text for details

The same magnetic mixing configuration has also been proved to better fit observables at low metallicities (in particular fluorine; [63]) and in Galactic open clusters [64], highlighting a sort of universality of the assumed free parameters (intrinsically linked to our still rather poor knowledge of magnetic field behaviour in stellar plasmas). In Fig. 3 we present a comparison between standard FRUITY models and new "magnetic" models for two metallicities ($Z = 0.01$ and $Z = 0.003$). With regard to the previous comparison, the differences are definitely greater. At $Z = 0.01$ (left panels), we observe an increase in element production with $30 < Z < 40$ and a decrease in surface abundances of elements heavier than Sr-Y-Zr. This fact is translated by a larger production of light s-only isotopes compared to the heaviest (such as $^{150}$Sm). For isotopes with $90 < A < 140$ we observe an increase up to a factor 2. Even more notable, the production of s-only isotopes in the region of the so-called weak s-process [32,65] results enhanced by a larger ratio (up to a factor 7). This could eventually make the contribution from AGB stars in this mass region comparable to that ascribed to massive stars. The same behaviour is confirmed at lower metallicities (right panels in Fig. 3), with the lightest s-only isotopes increasing by up to a factor of 10. From the point of view of absolute abundances, we note an increased surface production for elements up to $Z = 55$, while for the heavier ones we roughly confirm previous findings. We postpone a deeper analysis of these results to a future study [50], which would require the computation of a full set of AGB models to properly evaluate the effects on the GCE scale.

### 3.3 Comparison with recent evaluations of s- and r-residuals

Next we compare our GCE iterative model [43] with recent studies in the literature of the s- and/or r-components in the Solar system. The study of [66] estimates contributions from the r-process to Solar system abundances by adopting the waiting-point concept through a superposition of neutron density components normalised to the r-abundance peaks. They combine the results in early publications [67] with those in later improvements from experimental data and microscopic models [68,69]. The main advantage of this method is that it can exclude uncertainties relating to the modelling of the specific astrophysical sites for the production of the r-elements, which are still not very well known (see Sect. 1). The nuclear physics inputs for such calculations are understood only for the trans-Fe nuclei; therefore, their study is limited to the Sr–Pr region (atomic mass between A = 87–142). Then, they computed the s-component as $s = 1 - r$, just in the opposite way to the classical method.

Figure 4 compares the s-fractions at the Solar system formation epoch obtained by [43] (red dots) with those in [66] (blue open circles). The agreement is good (mean dispersion $\sim$ 7%) between both methods; [43] obtaining larger s-fractions in general. However, for a few nuclei, the difference is $>$ 25%. These nuclei are: $^{91,96}$Zr, $^{98}$Mo, $^{106}$Pd, and $^{118,122}$Sn. $^{91}$Zr is almost an s-only element according to [43], while $^{96}$Zr is almost an r-only isotope according to [66]. The rest of the discrepant nuclei have an intermediate nature between the s- and r-process according to both [43,66]. In fact, [66] also found a large difference in the s-process fraction for $^{98}$Mo, $^{106}$Pd, and $^{118}$Sn between the two different methods used by them (s = 1 − r and the stellar AGB model; see their Table 1). These authors note that in this mass range the cross-sections to the various (not always thermalised) isomeric states are almost unknown or very uncertain, as is the behaviour of the branching $\beta$-decays in the stellar plasma. This is an atomic mass zone where new experimental studies are needed. Among the $\beta$-decay rates to measure in highly ionised environments, the cases of $^{113,115}$Cd and $^{115}$In, possibly hosting non-thermalised isomeric states [54] are of particular interest. For the r-process, uncertainties may come in this mass range from estimates of the mass models of nuclei or from unreliable assumptions about the path followed at freeze out. In a number of cases $\beta$-delayed n-captures (i.e., last moment captures of remaining free neutrons) while the chain of decays from the progenitor nucleus to the valley of beta stability had already started, not considered in [66], may have an important role.

On the other hand, [70] used the latest release of the evaluated nuclear data file (ENDF/BVIII-0) library to obtain the solar system r-process abundances following the classical method. This library was first released in 1968 with its lat-est version in 2018 [71]. It incorporates data from the sixth edition of the atlas of neutron resonances. Neutron resonances dominate the astrophysical range of temperatures in many nuclei, and the comprehensive analysis of resonances is essential in nuclear astrophysics applications. The classical s-process calculation is done by [70] assuming $kT = 30$ keV and neutron fluence distribution parameters $f$ ($f$N($^{56}$Fe) being the amount of iron exposed to neutrons) and $\tau_o$ (neutron exposure) of 0.000402 $\pm$ 0.000083 and 0.3323 $\pm$ 0.0274, respectively. Although it is well known that the classical method approximation is not valid at the branching of the s-process, and that the s-process works in stars in a range of energies and neutron densities (in fact, the typical energy at which the main s-process occurs in stars is $kT \sim 8$ keV), we believe that it is a pedagogical exercise to compare our results with those of [70]. Figure 5 compares the abundances of the r-process at the Solar system epoch on the N(Si)$\equiv 10^6$ scale obtained in our iterative galactic model (red dots) with the corresponding results of [70] (open blue squares) and [66] (green dots). Comparison with [70] shows an important discrepancy (factors $\sim$ 1.5–5) in the mass range A = 70–80. The reason is that [43] included the contribution of the weak s-process coming from rotating massive stars [see also 35], which produces a significant amount of s-isotopes in that region and reduces the corresponding r-fractions. For the rest of the nuclei, the agreement is excellent (mean difference 11 $\pm$ 25%), with some discrepancies in the regions around A = 180 and A = 200, where there are several branches in the s-process path.

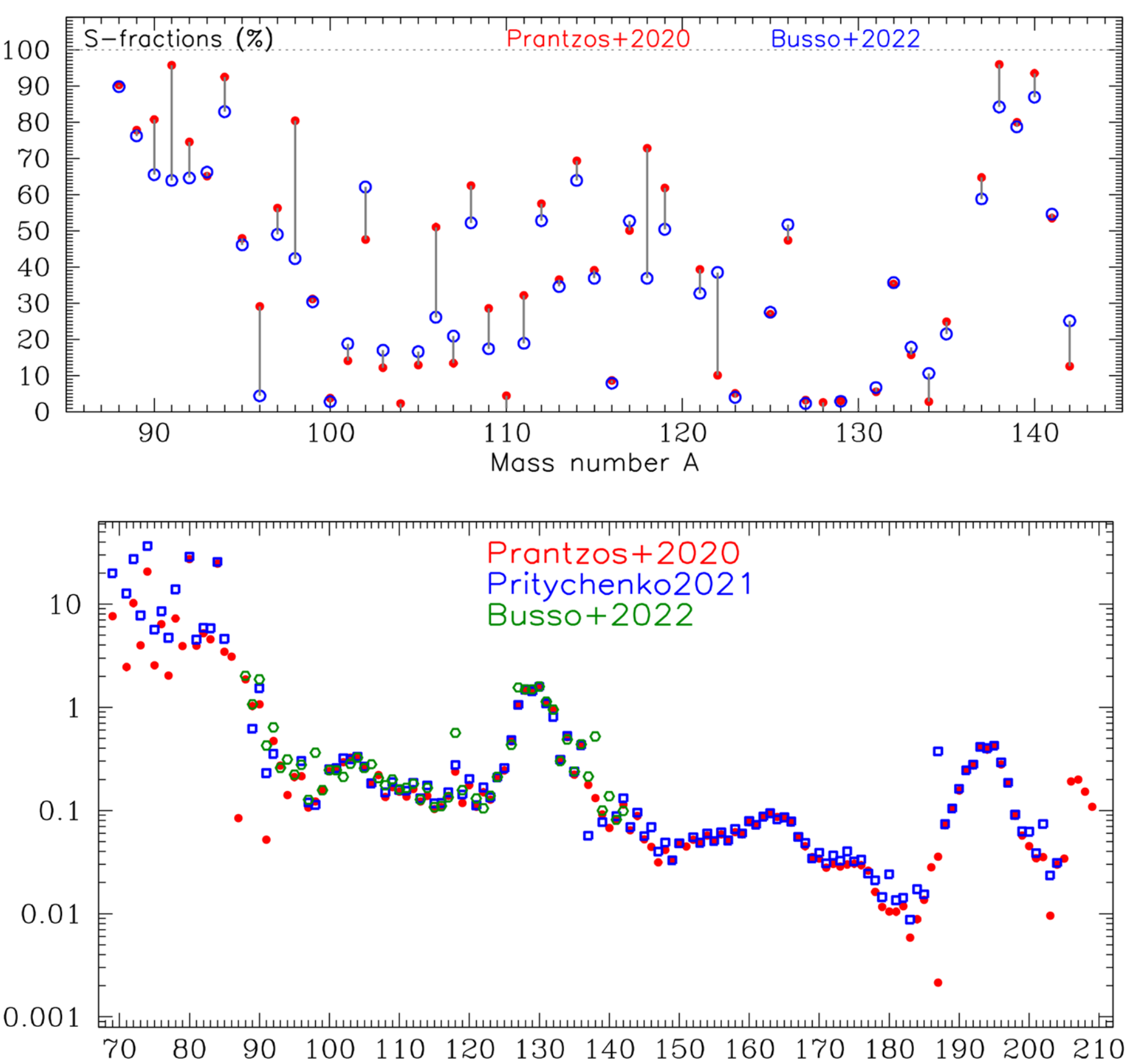

**Fig. 4** Solar s-process fractions (s-only nuclei excluded) of isotopes with mass number A between 87 and 142. Comparison between the evaluations of [43] (red filled circles) and [66] (blue open circles)

**Fig. 5** Solar r-process abundances in the scale of Si $= 10^6$ obtained by [43] (red), [70] (blue open squares) and [66] (green open circles); the latter are provided for isotopes with mass number between 87 and 142

### 3.4 Comparison to pre-solar SiC grains

In a recent study [72] molybdenum, ruthenium, and barium isotopes were simultaneously analysed in 55 individual presolar silicon carbide (SiC) grains from the Murchison CM2 meteorite. Most grains show clear s-process signatures, which are strongly correlated for molybdenum and ruthenium. Variations in s-process production observed for some nuclides reflect a strong dependence on the physical properties, neutron density, temperature, and timing, affecting various s-process branch points. For Mo and Ru, there is an overall good agreement between the grain data and the model calculations for AGB stars. According to the authors, their findings *" ... and the very good agreement with meteorite data seem to support a model without the need for a LEPP component to explain solar Mo and Ru*. This is in line with the conclusions of [27,35,43], also questioning the need for a LEPP.

In Fig. 6 we compare the s- and r- contributions to the pre-solar composition from the predictions of the stellar model of

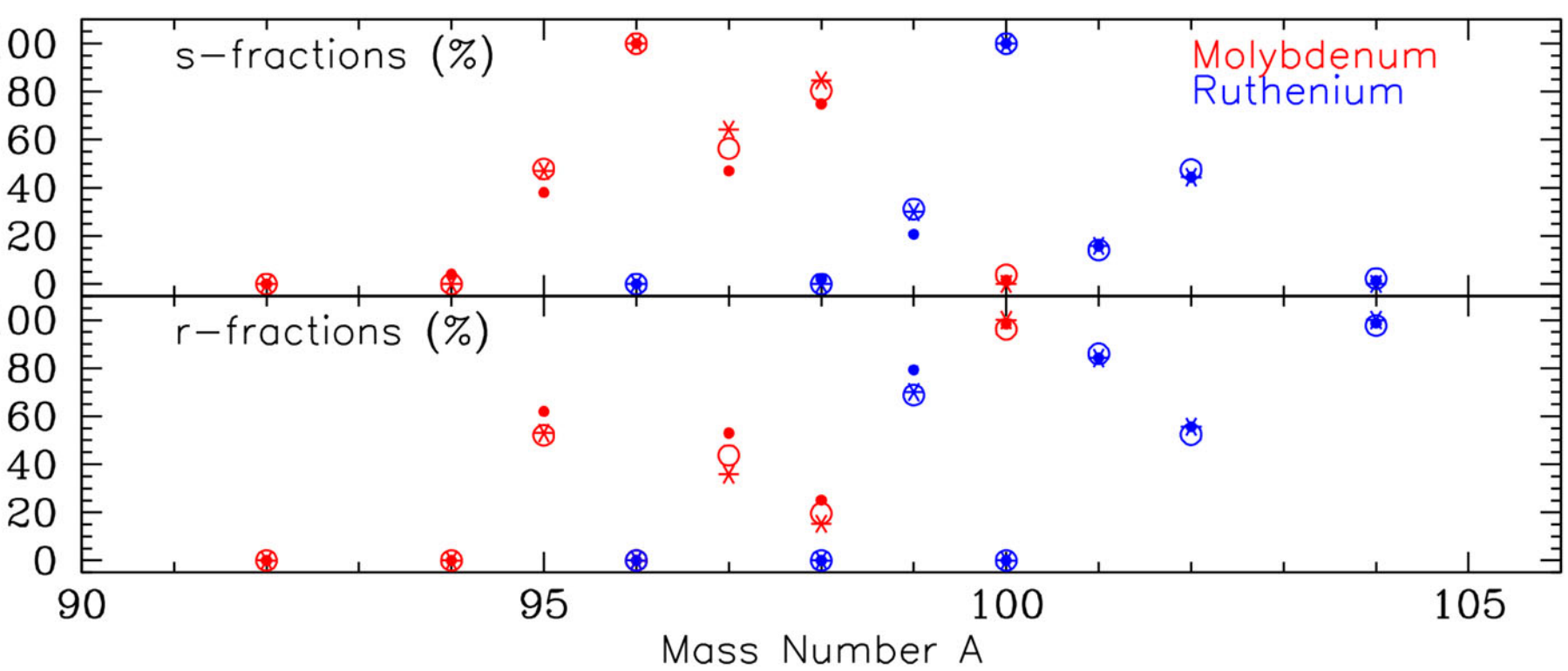

**Fig. 6** Solar s- and r- fractions for the isotopes of Mo (red) and Ru (blue) obtained by stellar models [11] (asterisks), iterative GCE [43] (open symbols), and [72] for presolar SiC grains (filled symbols)

[11] and the iterative GCE model of [43] to the grain results of [72]. The agreement is excellent between model and observations for the pure s- isotopes $^{96}$Mo and $^{100}$Ru, as well as for $^{101,102}$Ru and it is fair for $^{95,97,98}$Mo and $^{99}$Ru. The grain analysis finds a small contribution of the s-process, of the order of 1–2% , to $^{100}$Mo and $^{104}$Ru; this is also found in [43] through the rotating massive star contribution, but not in [11]. Finally, the grain analysis suggests a small contribution of the s-process (a few %) to $^{94}$Mo and $^{98}$Ru, while no s-contribution to those nuclides is found in the models of [11] and [43], where they are considered pure p-isotopes. This difference underscores the importance of the role of branchings in the s-process at $^{93}$Zr and $^{94}$Nb [72]. Other channels contributing to the synthesis of those nuclei, like the $\nu p$ process in hypernovae [73,74] may also play a significant role. More theoretical and observational work is needed to assess the importance of the aforementioned processes to the abundances of the Mo and Ru isotopes.

## 4 Summary

All the methods used up to now to derive the s- and r-process contributions have their own advantages and shortcomings. The earliest methods (Sects. 2.1 and 2.2) are characterised by simplicity, they help pinpointing nuclear uncertainties but they cannot encompass the variety of physical conditions (temperature, density, and metallicity variations) of the stellar sites of s-process. Stellar methods (Sect. 2.3) include by construction (some of) those physical conditions, but they face stellar modelling uncertainties on top of the nuclear ones (Sects. 3.1 and 3.2) and lack a consistent Galactic setting. GCE methods (Sects. 2.4 and 2.5) are made within a simple Galactic framework, and they face their own uncertainties – related to star formation, infall/outflow and stellar initial mass function (IMF) – on top of all the previous ones.

Regarding the GCE methods one should note that the resulting s-distribution at Solar system formation depends heavily on: (a) the IMF (which determines the "match" of the weak s-process in massive stars to the "main" s-process in AGBs),[3] and (b) even more importantly, on the details of the metallicity evolution, since the AGB yields are very sensitive to it and in different ways for the various ranges of atomic masses. One more complication arises from the fact that radial migration obviously played an important role in the evolution of the Galactic disc [see 76–78], implying that the Sun was formed a couple of kpc inward from its current position at 8 kpc. Its birth composition was then affected not only from the stars previously formed at its birth place but also from long-lived sources – AGB stars or neutron star mergers – that were formed elsewhere (in regions with different histories) and had time to migrate in that place and enrich it with their own s- and r-yields (see also [79] for the explanation of presolar grain origin). At present, there are no studies that evaluate the proto-solar s- and r-composition in such a complex Galactic environment. Some more time will be needed before the various uncertainties of the problem (nuclear, stellar, galactic) are reduced to a reasonable level.

**Acknowledgements** Part of this work was supported by the Spanish project PID2021-123110NB-I00 financed by MICIU/AEI/10.13039/501100011033 and by FEDER, una manera de hacer Europa, UE. CA dedicates this work to his all life partner Inma Domínguez, for her unconditional support and love throughout her life. Part of this work was supported by the European Union – NextGenerationEU RFF M4C2 1.1 PRIN 2022 project "2022RJLWHN URKA" and by INAF 2023 Theory Grant ObFu 1.05.23.06.06 "Understanding R-process and Kilonovae Aspects". SC dedicates this work to two individuals whose influence on him, both professionally and, most importantly, personally, has been profound: Roberto Gallino and Inma Domínguez.

**Data Availability Statement** Data will be made available on reasonable request. [Author's comment: The datasets generated during and/or analysed during the current study are available from the corresponding author on reasonable request.]

**Code Availability Statement** This manuscript has no associated code/software. [Author's comment: Code/Software sharing not applicable to this article as no code/software was generated or analysed during the current study.]

[3] It has recently been suggested [75] that a kind of secondary i- and s-process may develop in collapsar jets; it is hard at this point to evaluate their importance for the proto-solar contribution and the evaluation of the s- and r- residuals.